\documentclass[12pt]{article}
\usepackage{epsfig}
\usepackage{cite}

\begin{document}
\baselineskip 0.7cm

\newcommand{\gsim}{ \mathop{}_{\textstyle \sim}^{\textstyle >} }
\newcommand{\lsim}{ \mathop{}_{\textstyle \sim}^{\textstyle <} }
\newcommand{\vev}[1]{ \left\langle {#1} \right\rangle }
\newcommand{\lsp}{ \left ( }
\newcommand{\rsp}{ \right ) }
\newcommand{\lmp}{ \left \{ }
\newcommand{\rmp}{ \right \} }
\newcommand{\llp}{ \left [ }
\newcommand{\rlp}{ \right ] }
\newcommand{\labs}{ \left | }
\newcommand{\rabs}{ \right | }
\newcommand{\EV} { {\rm eV} }
\newcommand{\KEV}{ {\rm keV} }
\newcommand{\MEV}{ {\rm MeV} }
\newcommand{\GEV}{ {\rm GeV} }
\newcommand{\TEV}{ {\rm TeV} }
\newcommand{\YR}{ {\rm yr} }
\newcommand{\mgut}{M_{GUT}}
\newcommand{\mint}{M_{I}}
\newcommand{\mgra}{M_{3/2}}
\newcommand{\mll}{m_{\tilde{l}L}^{2}}
\newcommand{\mdr}{m_{\tilde{d}R}^{2}}
\newcommand{\mllXX}[1]{m_{\tilde{l}L , {#1}}^{2}}
\newcommand{\mdrXX}[1]{m_{\tilde{d}R , {#1}}^{2}}
\newcommand{\mgy}{m_{G1}}
\newcommand{\mgl}{m_{G2}}
\newcommand{\mgc}{m_{G3}}
\newcommand{\nuR}{\nu_{R}}
\newcommand{\slL}{\tilde{l}_{L}}
\newcommand{\slLi}{\tilde{l}_{Li}}
\newcommand{\sdR}{\tilde{d}_{R}}
\newcommand{\sdRi}{\tilde{d}_{Ri}}
\newcommand{\e}{{\rm e}}
\newcommand{\bsub}{\begin{subequations}}
\newcommand{\esub}{\end{subequations}}
\newcommand{\wt}{\widetilde}
\newcommand{\tm}{\times}
\newcommand{\ra}{\rightarrow}
\newcommand{\del}{\partial}
\newcommand{\az}{a_{Z}^{}}
\newcommand{\bz}{b_{Z}^{}}
\newcommand{\cz}{c_{Z}^{}}
\newcommand{\aw}{a_{W}^{}}
\newcommand{\bw}{b_{W}^{}}
\newcommand{\dw}{d_{W}^{}}
\newcommand{\sw}{s_{W}}
\newcommand{\cw}{c_{W}}
\newcommand{\gz}{g_{Z}^{}}
\newcommand{\mz}{m_{Z}^{}}
\newcommand{\pH}{p_{H}^{}}
\newcommand{\pone}{p_{1}^{}}
\newcommand{\ptwo}{p_{2}^{}}
\newcommand{\pt}{\partial}
\newcommand{\btable}{\begin{table}[htbp]\begin{center}}
\newcommand{\etable}[1]{ \end{tabular}\caption{#1}\end{center}\end{table} }
\newcommand{\vt}{\vspace{3mm}}
\renewcommand{\thefootnote}{\fnsymbol{footnote}}
\setcounter{footnote}{1}

\makeatletter
%
%
%
%
%
\newtoks\@stequation

\def\subequations{\refstepcounter{equation}%
  \edef\@savedequation{\the\c@equation}%
  \@stequation=\expandafter{\theequation}
  \edef\@savedtheequation{\the\@stequation}
  \edef\oldtheequation{\theequation}%
  \setcounter{equation}{0}%
  \def\theequation{\oldtheequation\alph{equation}}}

\def\endsubequations{%
  \ifnum\c@equation < 2 \@warning{Only \the\c@equation\space subequation
    used in equation \@savedequation}\fi
  \setcounter{equation}{\@savedequation}%
  \@stequation=\expandafter{\@savedtheequation}%
  \edef\theequation{\the\@stequation}%
  \global\@ignoretrue}


\def\eqnarray{\stepcounter{equation}\let\@currentlabel\theequation
\global\@eqnswtrue\m@th
\global\@eqcnt\z@\tabskip\@centering\let\\\@eqncr
$$\halign to\displaywidth\bgroup\@eqnsel\hskip\@centering
     $\displaystyle\tabskip\z@{##}$&\global\@eqcnt\@ne
      \hfil$\;{##}\;$\hfil
     &\global\@eqcnt\tw@ $\displaystyle\tabskip\z@{##}$\hfil
   \tabskip\@centering&\llap{##}\tabskip\z@\cr}

\makeatother


\begin{titlepage}

\begin{flushright}
UT-02-34
\end{flushright}

\vskip 0.35cm
\begin{center}
{\large \bf Neutrino Majorana Mass from Black Hole}

\vskip 0.4cm

Yosuke Uehara

\vskip 0.4cm

{\it Department of Physics, University of Tokyo, 
         Tokyo 113-0033, Japan}
\vskip 1.5cm

\abstract{We propose a new mechanism to generate the neutrino Majorana mass
in TeV-scale gravity models. The black hole violates
all non-gauged symmetries and can become the origin of lepton
number violating processes. The fluctuation of higher-dimensional
spacetime can result in the production of a black hole, which emits
2 neutrinos. If neutrinos are Majorana particles, this process is
equivalent to the free propagation of a neutrino with the insertion
of the black hole. From this fact, 
we derive the neutrino Majorana mass.
The result is completely consistent with the recently
observed evidence of neutrinoless double beta decay. And
the obtained neutrino Majorana mass satisfies 
the constraint from the density of the neutrino 
dark matter, which affects the cosmic structure formation.
Furthermore, we can explain the ultrahigh energy cosmic rays
by the Z-burst scenario with it.}

\end{center}
\end{titlepage}

\renewcommand{\thefootnote}{\arabic{footnote}}
\setcounter{footnote}{0}

%
%
%
%

TeV-scale gravity models propose that the true fundamental
scale is $\mbox{O}(\TEV)$, and thus many effects of the
quantum gravity will appear in future experiments. The most appealing
phenomenon is the production of TeV-scale black holes. $\TEV$-scale
colliders or high-energy cosmic ray detectors can be the candidates
of black hole factories.
The black hole is considered to break all non-gauged
symmetries. So we can expect the lepton number violating
processes occur in the decay of black holes \cite{LC}. 

In this letter, we consider a new mechanism to generate
the neutrino Majorana mass from a black hole. The existence
of the neutrino Majorana mass implies that the lepton number is violated
in nature. \cite{Doublebeta} announced that the HEIDELBERG-
MOSCOW double beta decay experiment observed the evidence for
neutrinoless double beta decay. It will be the first
evidence of the lepton number violation in the world, if confirmed.
Our mechanism naturally explains the result of \cite{Doublebeta}.
The obtained Majorana mass is consistent with the expected density of 
the neutrino dark matter in the universe.
Furthermore its value enables the Z-burst scenario
to explain cosmic rays above the GZK cutoff.

In TeV-scale gravity models, we have to identify the number of
extra dimensions $n$, and the true fundamental scale $M_{D}$.
Since superstring theory or M-theory suggest 10-dimensional or 
11-dimensional spacetime, we consider the cases of $n=6,7$ in this letter.

As is pointed out in \cite{Giddings,Dimopoulos-Landsberg}, the
decay of black holes do not discriminate any of the Standard
Model (SM) particles. The possibility that a certain particle
being emitted from a black hole depends on the degree of freedom
of the particle. That of the SM is about 120, of a neutrino
is 2, and of a Higgs boson is 1.

Since we do not know the fundamental theory of the quantum gravity,
we have to depend on the semiclassical approximation. To validate this 
description, the entropy of a black hole must be large
enough \cite{Cheung}. From this requirement, we have $M_{BH} \gsim 5 M_{D}$,
where $M_{BH}$ is the black hole mass. And as we raise the 
black hole mass, the production rate of the black hole 
is reduced drastically. So we
consider a black hole with mass $M_{BH} \sim 5 M_{D}$.

From figure \ref{Nfig}, we observe that the number of
particles emitted from a black hole is about 4
if $M_{BH} \sim 5 M_{D}$. So we can consider
the production of 2 neutrinos and 2 Higgs bosons from
a black hole which is produced by the fluctuation of 
(n+4)-dimensional spacetime.

Let us assume that neutrinos are Majorana particles.
Then this process leads to the Majorana mass term of neutrinos, as
can be seen in figure \ref{BHfig}. Since $\nu^{c}$ is the antiparticle
of $\nu$, figure \ref{BHfig} can be interpreted as the propagation
of $\nu$, with the insertion of a black hole and 2 Higgs bosons.
This insertion leads to the neutrino Majorana mass.

\begin{figure}[htbp]
\begin{center}
\begin{minipage}{5cm}
\begin{center}
\centerline{\psfig{file=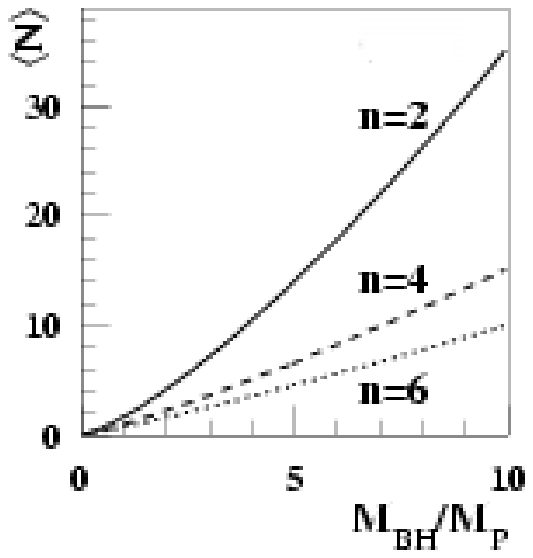,height=6cm}}
\caption{The average decay multiplicity for a Schwarzschild black hole
 \cite{Dimopoulos-Landsberg}. Here $M_{P}=M_{D}$.}
\label{Nfig}
\end{center}
\end{minipage}
\hspace{3cm}
\begin{minipage}{5cm}
\begin{center}
\centerline{\psfig{file=BHfig.eps,height=6cm}}
\caption{The Feynman diagram responsible for the Majorana mass term of neutrinos.}
\label{BHfig}
\end{center}
\end{minipage}
\end{center}
\end{figure}

Now we estimate the neutrino Majorana mass obtained in the process
figure \ref{BHfig}.

Firstly, the black hole must survive
until they decay into 2 neutrinos. The Planck time
of $(n+4)$-dimensional spacetime is given by
\cite{Giudice-Rattazzi-Wells}:
\begin{eqnarray}
t_{pl} = \left[ \frac{(2 \pi)^{n-1}}{4 M_{D}^{n+2}} \right]^{1/(n+2)},
\end{eqnarray}
and from the uncertainty principle, we can violate the conservation
low of energy during $\Delta t$. It is:
\begin{eqnarray}
\Delta t = \frac{1}{\Delta E} = \frac{1}{5 M_{D}}.
\end{eqnarray}
So the suppression factor arises from the time instability of
the black hole is:
\begin{eqnarray}
\frac{\Delta t}{t_{pl}} \sim \frac{1}{15}.
\end{eqnarray}

Secondly, the black hole should hold the energy $M_{BH} \sim  5 M_{D}$ 
inside the $(n+4)$-dimensional Schwarzshild black hole, whose radius is
$R_{S}$. This fact leads to the suppression factor:
\begin{eqnarray} 
\left( \frac{1}{5 M_{D} R_{S}} \right)^{3} \sim (\frac{1}{5})^{3},
\end{eqnarray}
where $R_{S}$ is given by \cite{Myers-Perry}:
\begin{eqnarray}
R_{S} = \frac{1}{\sqrt{\pi} M_{D}} \left[ \frac{M_{BH}}{M_{D}} \left( \frac{8 \Gamma(\frac{n+3}{2})}{n+2} \right)\right]^{1/(n+1)}.
\end{eqnarray}

Thirdly, the black hole must emit 2 neutrinos
(since these neutrinos form the Majorana mass term, their
degree of freedom is 2.) and 2 Higgs bosons
which acquire the vacuum expectation values (VEV). 
The probability of this decay mode being realized is given by:
\begin{eqnarray}
\frac{2 \cdot 1^{2}}{121^{4}} \sim 9.3 \tm 10^{-9}.
\end{eqnarray}

Finally the VEV of Higgs bosons $v$ is suppressed by the
true fundamental scale $M_{D}$. So the existence of 2 Higgs bosons 
results in the suppression factor:
\begin{eqnarray}
(\frac{v}{M_{D}})^{2} = (\frac{0.174 \ \TEV}{M_{D}})^{2}.
\end{eqnarray}

Including all of the results, the neutrino Majorana mass becomes:
\begin{eqnarray}
m &=& M_{BH} \ (\frac{1}{15}) \ (\frac{1}{5})^{3} \ (9.3 \tm 10^{-9}) \ (\frac{0.174 \ \TEV}{M_{D}})^{2} \nonumber \\
&=& 0.75 \ \EV \ (\frac{1 \ \TEV}{M_{D}}).
\end{eqnarray}

Here we obtain the tiny neutrino Majorana mass using the 
arguments about the TeV-scale black hole only. From now
we discuss the obtained neutrino Majorana mass with experimental results.

Consider about neutrinoless double beta decay.
\cite{Doublebeta} searched the decay mode
$^{76} Ge \ra ^{76} Se \ + \ 2 e^{-}$, and reported the half
life to be:
\begin{eqnarray}
T_{1/2}^{0 \nu} = (0.8 - 18.3) \tm 10^{25} \ \mbox{yr}.
\end{eqnarray}
This result can be interpreted as the existence of neutrino effective mass:
\begin{eqnarray}
\vev{m} = (0.11 - 0.56) \ \EV \ \mbox{with best fit} \ 0.39 \ \EV, \label{expm}
\end{eqnarray}
which is defined by $\vev{m} \equiv |m_{1} |U_{e1}|^{2} + m_{2}
|U_{e2}|^{2} + m_{3} |U_{e3}|^{2}|$. Here $U_{\alpha i}$ are the MNS
matrix elements \cite{MNS}, and for the simplicity we do not consider
about CP-violation in the lepton sector now. We justify
this neglectfulness later.

In our setup, neutrinoless double beta decay is induced by a black
hole. Its explicit process is shown in figure \ref{BHdoublefig}.
\begin{figure}[htbp]
\begin{center}
\centerline{\psfig{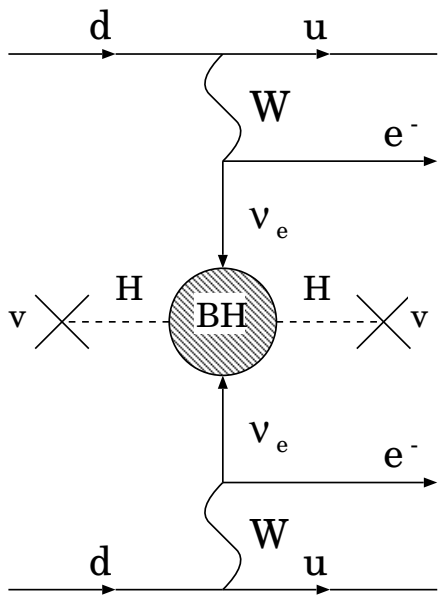}}
\caption{The explicit example of neutrinoless double beta decay induced
by a black hole.}
\label{BHdoublefig}
\end{center}
\end{figure}
\vspace{-1cm}

The neutrino Majorana mass is generated by a black hole,
resulting in $m = (0.75 \ \EV) \ (1 \ \TEV)/(M_{D})$. So 3 generation
neutrinos are almost degenerate, and radiative corrections split their
masses. From CHOOZ experiment \cite{CHOOZ}
we have $|U_{e3}|^{2} < 0.03$ and we neglect this term. Then
$\vev{m}$ is roughly given by $2 m$. From equation (\ref{expm}),
in order to explain the observed neutrinoless double beta decay 
by our mechanism, we obtain:
\begin{eqnarray}
0.11 \le 1.5 \frac{1 \ \TEV}{M_{D}} \le 0.56, \ \mbox{with best fit} \ 0.39
\end{eqnarray}
This result can be interpreted as a constraint on $M_{D}$. It becomes:
\begin{eqnarray}
2.7 \TEV \ \le M_{D} \ \le 14 \TEV, \ \mbox{with best fit} \ 3.8 \ \TEV. \label{reseq}
\end{eqnarray}
From naturalness of TeV-scale gravity models, $M_{D}$ should not
exceed $1 \TEV$ too much. But as you can see from equation
(\ref{reseq}), we can explain the result of \cite{Doublebeta}
without spoiling it. If we take into account the effects of
CP-violation, $\vev{m}$ becomes smaller because a CP-violating
phase cancels the contribution between $m_{1}$ and $m_{2}$.
As a result, the value of $M_{D}$ is also reduced and 
naturalness improves. So the neglectfulness of CP-violation
do not affect our discussions.

So we arrive at an interesting result. {\bf The TeV-scale black hole can
be a very natural origin of the observed neutrinoless double beta decay.}
If a CP-violating phase modifies the allowed region of $M_{D}$,
There exists a possibility that the LHC can observe semiclassical
TeV-scale black holes. 
In that case we can compare the value of $M_{D}$ obtained
from neutrinoless double beta decay experiments and at the LHC,
and verify whether the neutrino Majorana mass is generated by
a black hole or not.

Hereafter we consider about the astrophysical constraints on the masses of
neutrinos, and show you that our results are consistent with them.

First consider about the dark matter in the universe. Neutrinos
can be the candidates of the hot dark matter, and their masses affect
the evolution of the universe.
The most stringent constraint in the case of three degenerate neutrinos is 
obtained by the considerations on the cosmic structure formation
in the low-matter density universe. It becomes \cite{Fukugita}:
\begin{eqnarray}
\sum_{i=1}^{3} \le 1.8 \ \EV,
\end{eqnarray}
where $i$ denotes the generation of neutrino. 
To satisfy this constraint, the true fundamental scale $M_{D}$ should be:
\begin{eqnarray}
M_{D} \ge 1.3 \ \TEV. \label{DMeq}
\end{eqnarray}
This is consistent with equation (\ref{reseq}).
And from the identity:
$\Omega_{\nu} h^{2} = \sum_{i=1}^{3} m_{i}/(93.8 \ \EV)$,
we obtain:
\begin{eqnarray}
\Omega_{\nu} h^{2} = 0.024 \frac{1 \ \TEV}{M_{D}} \le 0.018.
\end{eqnarray}
So the neutrino component can contribute to the matter density
of the universe less than about $13 \%$. (Here we assumed that
$h=0.7$ \cite{Freedman} and $\Omega_{m}=0.28$ \cite{Perlmutter,Riess}.)

Next consider about the ultrahigh energy cosmic rays.
After the measurement of the cosmic microwave background (CMB) radiation,
it was claimed that the spectra of cosmic rays suddenly dump
at the energy $E \sim 4 \tm 10^{19} \ \EV$ (the GZK cutoff), since at the energy
cosmic rays result in the interaction with the CMB photons
and lose their energies \cite{GZK}. 
But some experiments like AGASA \cite{AGASA} found cosmic rays above the
cutoff. This is a serious challange not only for astrophysicists 
but also for particle physicists.

Many scenarios are proposed to explain such cosmic rays, 
but here we consider about the Z-burst
scenario. It is based on the Z boson production resulting from
the resonant annihilation of ultrahigh energy cosmic neutrinos
with relic neutrinos into Z bosons. Since neutrinos do not
interact with the CMB photon even if their energies are very high, they
can travel a very long distance. So we can explain cosmic rays above the
GZK cutoff if they annihilate into Z boson near our galaxy, escaping
from the short attenuation length of ultrahigh energy cosmic rays.

\cite{Fodor} showed that in the case ordinary cosmic rays
are protons of extragalactic origin, the required neutrino mass
to explain current experiments is given by:
\begin{eqnarray}
0.08 \EV \ \le m_{\nu}\  \le 1.3 \EV.
\end{eqnarray}
Our results can satisfy this condition if:
\begin{eqnarray}
0.58 \TEV \ \le M_{D} \ \le 9.4 \TEV.
\end{eqnarray}
From equation (\ref{DMeq}), we cannot explain some region of
this constraint. But most of the region are consistent with the
previous results.

Now we show you the three obtained constraints on the true fundamental
scale $M_{D}$ in our model, namely the neutrino Majorana mass is 
generated by the TeV-scale black hole.
\bsub
\begin{eqnarray}
&2.7& \TEV \le M_{D} \le 14 \ \TEV \ \mbox{with best fit} \ M_{D}=3.8 \ \TEV, \nonumber \\
&&\mbox{in order to explain the observed neutrinoless double beta decay.} \\
&M_{D}& \ge 1.3 \ \TEV, \nonumber \\
&&\mbox{from the considerations about the density of the neutrino dark matter.} \nonumber \\
&& \\
&0.58& \TEV \le M_{D} \le 9.4 \ \TEV, \nonumber \\
&& \mbox{in order to explain cosmic rays above the GZK cutoff by the Z-burst scenario.} \nonumber \\
&& 
\end{eqnarray}
\esub

So we can explain three experimental results stated above by assuming:
\begin{eqnarray}
2.7 \ \TEV \le M_{D} \le 9.4 \ \TEV.
\end{eqnarray}

To summarize, we propose a new mechanism to generate the neutrino
Majorana mass in the context of TeV-scale gravity models.
The TeV-scale black hole can generate the neutrino Majorana mass, 
and the observed neutrinoless double beta decay
is naturally explained by the mechanism. If it is correct,
the true fundamental scale $M_{D}$ measured by neutrinoless double beta decay
experiments and at the future colliders should agree. This can be
a definite test of the results shown in this letter.
The obtained neutrino Majorana mass is consistent 
with the astrophysical constraint
on the density of the neutrino dark matter, which affects the cosmic structure 
formation. And it enables us to explain cosmic rays above the GZK
cutoff by the Z-burst scenario.

\vt

{\bf Acknowledgment}

Y.U. thank Japan Society for the Promotion of Science for financial
support.

\vt

\end{document}